\documentclass{article}

\usepackage{arxiv}

\usepackage[utf8]{inputenc} % allow utf-8 input
\usepackage[T1]{fontenc}    % use 8-bit T1 fonts
\usepackage{hyperref}       % hyperlinks
\usepackage{url}            % simple URL typesetting
\usepackage{booktabs}       % professional-quality tables
\usepackage{amsfonts}       % blackboard math symbols
\usepackage{nicefrac}       % compact symbols for 1/2, etc.
\usepackage{microtype}      % microtypography
\usepackage{lipsum}
\usepackage{graphicx}
\usepackage{threeparttable}
\usepackage{makecell}
\usepackage{diagbox}
\usepackage{xurl}

\title{Self-Sovereign Identity for Consented and Content-Based Access to Medical Records using Blockchain}

\author{
Marie Tcholakian \\
  Samovar, Télécom SudParis \\
  Institut Polytechnique de Paris \\
  France, Palaiseau 91120 \\
  \texttt{}
    \And
Karolina Gorna \\
  Samovar, Télécom SudParis \\
  Institut Polytechnique de Paris \\
  France, Palaiseau 91120 \\
  \texttt{}
   \And
 Maryline Laurent* \\
  Samovar, Télécom SudParis \\
  Institut Polytechnique de Paris \\
  France, Palaiseau 91120 \\
  \texttt{} \\
  \And
  Hella Kaffel Ben Ayed \\
   Faculty of Science of Tunis \\
   University of Tunis El Manar \\
   Tunisia, El Manar Tunis 2092 \\\
   \texttt{}
   \And
   Montassar Naghmouchi \\
     Samovar, Télécom SudParis \\
  Institut Polytechnique de Paris \\
  France, Palaiseau 91120 \\
  \texttt{} \\}
\begin{document}
\maketitle
\begin{abstract}
Electronic Health Records (EHRs) and Medical Data are classified as personal data in every privacy law, meaning that any related service that includes processing such data must come with full security, confidentiality, privacy and accountability. Solutions for health data management, as in storing it, sharing and processing it, are emerging quickly and were significantly boosted by the Covid-19 pandemic that created a need to move things online. 
EHRs makes a crucial part of digital identity data, and the same digital identity trends – as in self sovereign identity powered by decentralized ledger technologies like Blockchain, are being researched or implemented in contexts managing digital interactions between health facilities, patients and health professionals.
In this paper, we propose a blockchain-based solution enabling secure exchange of EHRs between different parties powered by a self-sovereign identity (SSI) wallet and decentralized identifiers. We also make use of a consortium IPFS network for off-chain storage and attribute-based encryption (ABE) to ensure data confidentiality and integrity.
Through our solution, we grant users full control over their medical data, and enable them to securely share it in total confidentiality over secure communication channels between user wallets using encryption. We also use DIDs for better user privacy and limit any possible correlations or identification by using pairwise DIDs. Overall, combining this set of technologies guarantees secure exchange of EHRs, secure storage and management along with by-design features inherited from the technological stack.
\end{abstract}
\textbf{\textit{Abbreviations}} \\
\textbf{ABE}: Attribute Based Encryption; \textbf{EHR}: Electronic Health Record; \textbf{AES}: Advanced Encryption Standard; \\
\textbf{CID}: Content ID; \textbf{CP-ABE}: Ciphertext-Policy Attribute Based Encryption; \textbf{DID}: Decentralized Identifier, \\
\textbf{IPFS}: InterPlanetary File System; \textbf{SSI}: Self-sovereign identity.

\textbf{\textit{Correspondence}} should be addressed to 
\textit{maryline.laurent@telecom-sudparis.eu} (Maryline Laurent)
\bigskip

\section{Introduction}
\label{Intro}
EHRs and health related data have always been of interest to hackers due to their personal private nature, and Covid-19 was a landmark around the world in terms of health data collection, leading to more and more serious attacks. For instance, HIPAA reports healthcare data breaches \cite{hipa} in the US on medical records that were reported to the US HHS’ Oﬃce for Civil Rights (OCR) in
January 2022. They observed an increase by 38.9\% of healthcare data breaches in January 2022 compared to January 2020. These breaches aﬀect thousands of records and millions of patients. Most of these breaches occur at the network servers of healthcare providers. Ransomware, phishing, and unauthorized access are the causes of healthcare data breaches in January 2022. According to the studied breaches and other resources \cite{Papaioannou}, we can identify the following threats targeting health data : 1) Impersonation where the attacker pretends to be legitimate to gain access to medical data reference. This can compromise the confidentiality and integrity of data. This may also bring into question the liability of the healthcare professionals with respect to the access and authorizations of which they are the object, 2) Malicious code injection attack may result in modifying the stored data which compromises medical data integrity, and 3) Authentication and Identity-based attacks that are among the most dangerous attacks on patient data. These threats target the authentication process allowing malicious users to be authenticated and afterward to transmit fake data.

To prevent these threats, the following functional and non-functional security requirements are identified \cite{Papaioannou}:
\begin{itemize}
\item Data confidentiality and integrity as medical records are considered confidential and tamper-proofed throughout their whole life cycle I.e., generation, storage, transmission, and processing.
\item Accountability and non-repudiation to prevent participating entities from denying previous commitments or actions related to data processing.
\item Strong identification via unique, global and permanent identifiers that can enable strong and secure authentication with a high level of assurance (LoA).
\item Access control to provide restricted access to the medical records according to the requester’s authorizations.
\end{itemize}
Traditional centralized access control systems suﬀer from single points of failure as their centralized servers can be unreachable in case of attacks, or lack of connectivity. The traditional PKI-based authentication solutions are inefficient. The level of complexity of the certificate path processing in a healthcare PKI infrastructure is one factor that aﬀects the efficient adoption of the PKI technology in healthcare networks. Furthermore, they do not support control by the users \cite{Shuaib}. Decentralization is then required to overcome the disadvantages and limitations of the existing centralized cloud-based healthcare systems. Despite the eﬀorts on access control mechanisms for medical data and auditability in the digital healthcare ecosystem, there are still many open issues to address for the development of robust and user-centric access control mechanisms \cite{Papaioannou}.

New models and mechanisms of digital identification, authentication and access control are needed for a healthcare ecosystem that is decentralized by nature and made of multiple stakeholders and has high requirements in terms of compliance to regulations and reliability.

The Self-Sovereign Identity model, SSI,  deﬁnes a new approach to create and manage digital decentralized identity via blockchain-based identifiers and verifiable claims. It is a user-centric model that comes with less dependency on identity providers by allowing the users to register themselves and obtain controllable identifiers called DIDs that can be linked to claims in the form of a Verifiable Credential issued by an issuer. These credentials are fully controlled by the users and they are verifiable via blockchain without relying on the issuer, moreover they come with different privacy aware methods like zero-knowledge proofs and selective disclosure.
A digital movement that recognizes that individuals should own and control their digital identity without relying on a third party is built around the model and many communities are being established studying different possible use-cases with different ambitions that range from simple identity wallets to building a full decentralized identity layer for the internet. 
Some studies on SSI in the healthcare sector are emerging. However, to the best of our knowledge, they are limited to surveys and prospective studies.

This paper addresses the access management of health records during their life-cycle. We address the problems of patient consent and authorizing access to their data, as well as the accountability of healthcare professionals and institutions that were granted access to this data. The access to the medical records is strengthened through a content-based access control
ensured with a two level encryption scheme: symmetric AES encryption for privately storing medical records into the Inter-Planetary File System (IPFS), and Attribute-Based Encryption (ABE) of AES keys where the access policy is written into the ABE cipher-text. As such, the consent by the user for aﬀording access to the medical staﬀ is translated into the patient asking for modifying the ciphertext policy, and each given consent is logged into the blockchain. The proposed SSI architecture is scalable and relies on an ID wallet which is provided to each patient and medical staﬀ. The wallet embeds the attributes of the patients, e.g. name, surname, social security ID, payment features, or of the doctor, e.g. name, surname, oﬃcial license number, hospital patient ID, doctor’s medical department... These attributes are endorsed by the national health authority or the hospital itself.

The remainder of this paper is organized as follows: Section 2 summarizes the literature review and the related works to our addressed problem. Section 3 gives the background helpful for understanding the concepts and technologies  underlying our solution. Section 4 presents our proposed architecture as well as our reference scenario, and in Section 5 discusses our architecture. 
Finally in  Section 6 gives conclusions of our work.
\section{Related Work}
\label{Related}

SSI systems rely on DLTs and the most common ones are based on a public or consortium blockchain network. Blockchain comes with built-in features like permanent tamper-proof transactions, decentralization and shared governance, and can be used to ensure the integrity of data. Many pilot projects based on blockchain technology are underway in various domains where security, trust and reliability of transactions among various entities are required \cite{Alam}. Healthcare is one of these domains. In healthcare, Blockchain is used for maintaining and exchanging medical records and for management of the medical supply chain \cite{Krawiec}. In \cite{Krawiec}, an illustrative healthcare Blockchain ecosystem architecture is presented. The question regarding the storage of medical information either “On-chain” or 
“Oﬀ-chain” is discussed considering the security, the availability and the performance features. Smart contracts are proposed as a mechanism to enforce a standardized data submission for blockchain transactions, enabling blockchain to act as an interoperable transaction layer for nationwide health systems either by storing on-chain publicly accessible data or by storing pointers for off-chain privately stored data on a given database.
Various applications of Blockchain in healthcare are emphasized in \cite{Alam}: 1) Health data exchange in a secure and reliable manner, 2) Sharing EHR for research purposes while maintaining subject/patient anonymity, 3) EHR interoperability for cooperation between various entities, 4) Efficient health insurance claim processing, and 5) Efficient and reliable drug and medical equipment supply.
The paper \cite{Siqueira} is a comprehensive systematic survey on the use of Blockchain technology and SSI in healthcare. The survey shows that Blockchain is a suitable alternative for EHR management. It facilitates users’ access to their health records. Although using Blockchain to manage health data can prevent data tampering, it cannot address entirely the privacy issues \cite{Alam}\cite{Shuaib}. Furthermore, more mechanisms are required to empower patients by giving them full control over their EHR using SSI systems. \cite{Shuaib} identiﬁes the following requirements for  the adoption of SSI in the ﬁeld of healthcare: trust (integrity and control), transparency (no secret transmission of knowledge), ease of use, key recovery, security (only authorized access), access (maintenance, correction, and auditability), compliance with regulation, efficiency (no redundancy) and patient awareness (consensual private data sharing). The authors discuss the factors of SSI-based healthcare from the perspective of the stakeholders’ needs.

\section{Background}
\label{Background}
\subsection{Self-Sovereign Identity (SSI)}
\begin{figure}[!h]
    \centering
    \includegraphics[scale=0.3]{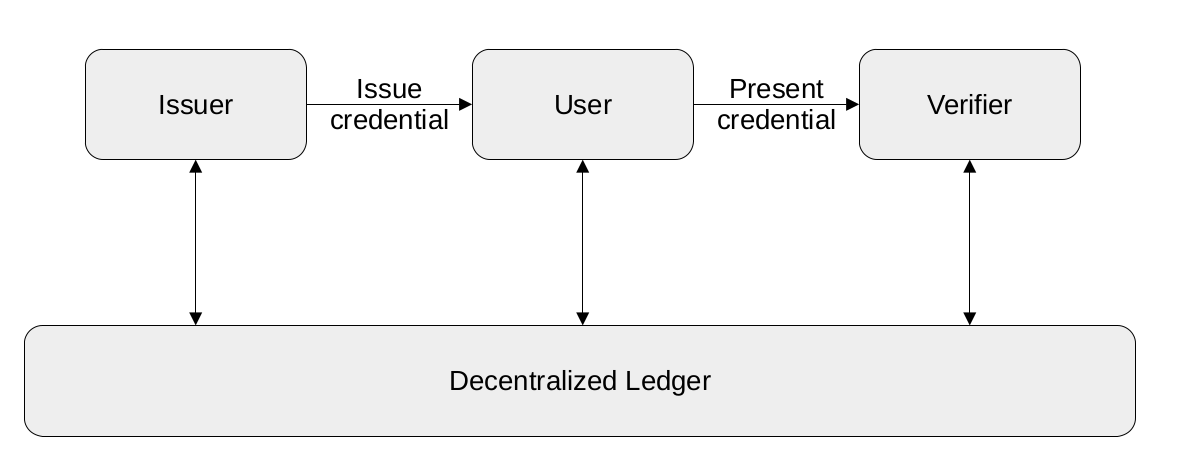}
    \caption{Self-Sovereign Identity model, based on a decentralized ledger}
    \label{fig:SSI}    
\end{figure}

SSI concepts and principles evolved through multiple identity workshops and papers, starting from the paper “laws of identity” by Kim Cameron in 2005 to the blog post of Christopher Allen \cite{path-to-ssi} in 2016 that coined the term SSI and set 10 principles for an SSI system. Supported by the Web of Trust Community \cite{git-weboftrust}, these principles are:
\begin{itemize}
    \item \textbf{Existence:} Users must have an independent autonomous existence.
    \item \textbf{Control:} Users must control their identities.
    \item \textbf{Access:} Users must have access to their own data.
    \item \textbf{Transparency:} Systems and algorithms must be transparent.
    \item \textbf{Persistence:} Identities must be long-lived.
    \item \textbf{Portability:} Information and services about identity must be transportable.
    \item \textbf{Interoperability:} Identities should be as widely usable as possible.
    \item \textbf{Consent:} Users must agree to the use of their identity.
    \item \textbf{Minimization:} Disclosure of claims must be minimized.
    \item \textbf{Protection:} The rights of users must be protected.
\end{itemize} 
SSI model, depicted in Figure \ref{fig:SSI}, leverages user autonomy by eliminating registration authorities, and gives users full control over their identifiers and authentication material. Moreover, SSI eliminates the role of Identity Providers that are very influential in other identity models like the federated model and the user centric model \cite{Dig-Identity-Management}. 

Beyond that, SSI enables to certify the authenticity of identities and personal identifiable information (known as attributes, e.g. date of birth, citizenship or university degrees) to service providers as some authorities are assumed to endorse identities and subset of attributes to users under the form of verifiable claims. Identities, attributes and verifiable claims are kept in a safe place within the digital wallet of their owner. 
An ID wallet holder can use his digital wallet, to authenticate or prove some attribute ownership with the credentials he has been issued by ID authorities, e.g. national health authority for social security number, prefecture for national ID and driving license. The holder is identified in the
whole system using a global Decentralized Identiﬁer, like the W3C DID standard.
\subsection{Decentralized Identifiers: DID standard}
\begin{figure}[!h]
    \centering
    \includegraphics[scale=0.3]{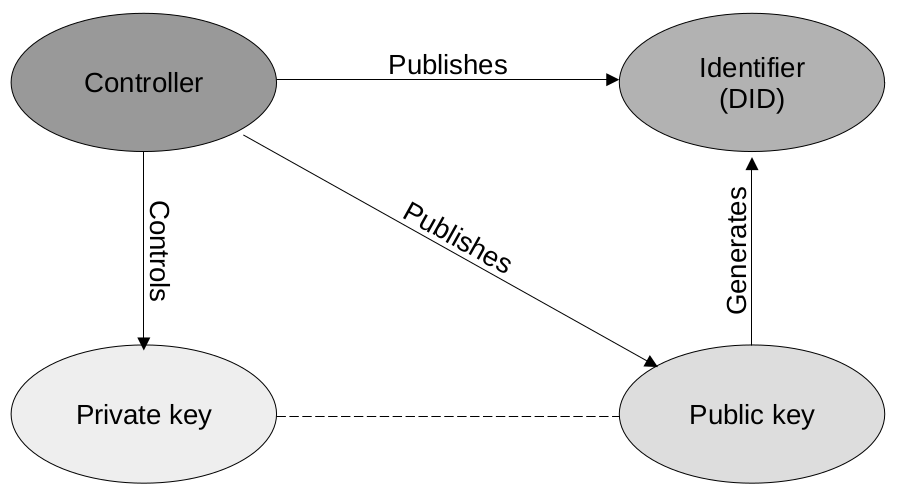}
    \caption{Relationships between DID, Public keys and private keys \cite{ssimeetup}}
    \label{fig:DID}    
\end{figure}

Decentralized Identifiers, DID as an acronym, is a W3C standard that defines identifiers that satisfy the ten requirements for SSI systems. DIDs are decentralized, meaning that no centralized authorities are needed even for the registration, giving users full autonomy in registering and using their identifiers. Moreover, a DID is unique and permanent and directly controllable by its owner or controller, since DID standard supports delegation. Using public key cryptography, a DID is linked to a public key, meaning that the DID controller can cryptographically prove that they control the DID, allowing for authentication and more importantly: linking the DID to a set of claims in the form of a Verifiable Credential. This means that DID are the core component of SSI systems that are based on the exchange of Credentials and enabled by a public or consortium DLT.
Figure \ref{fig:DID}  depicts the relationship between a DID and private and public keys.
Apart from that, a DID is discoverable and resolvable, meaning that we can reach out to the owner or controller for different interactions. DIDs are also used to create secure communication channels after mutual authentication and can be suited for private use by introducing pairwise DIDs that are used between two and only two parties, unlike anywise DIDs that are global and public and generally published on the ledger. A pairwise DID, or any other N-wise DID, is only resolvable by the designated entities, unlike anywise DIDs that are publicly resolvable \cite{peer-DID}.

These capacities make DIDs a very powerful standard that can empower true SSI platforms.

\subsection{Attribute-Based Encryption}
\label{ABE}
Attribute-Based Encryption (ABE) is an asymmetric encryption algorithm in which the secret key of a user and the ciphertext are relying on attributes. The users are characterized with a set of attributes, e.g. country of residence, a profession... For each of them, the users are assigned a secret which can be used to prove the validity of the related attributes. Decryption of ciphertext is only possible if there is a match between the attributes owned by the user and the attributes considered for the ciphertext. This ABE scheme builds on an authority which is in charge of generating users' secrets from its master key. Four algorithms are needed to satisfy the ABE scheme: 
\begin{enumerate}
    \item \textbf{Set up} takes as inputs a security parameter and a set of attributes $\Omega$ and outputs a master key $msk$ and some public parameters, e.g. a public key $pk$.
    \item \textbf{Key generation} which, given a set of attributes $A$ and the master key $msk$, generates a set of related secret keys $sk_A$.
    \item \textbf{Encryption} which, given a message $M$, the public key $pk$ and a policy $\phi$ (e.g. a Boolean formula) produces a ciphertext $C_\phi$.
    \item \textbf{Decryption} which, from a secret key $sk_A$ and an encrypted message $C_\phi$, outputs either the message $M$ if $\phi(A)$ satisfies the policy $\phi$ or an error. 
\end{enumerate}

\begin{figure}[!h]
    \centering
    \includegraphics[scale=0.3]{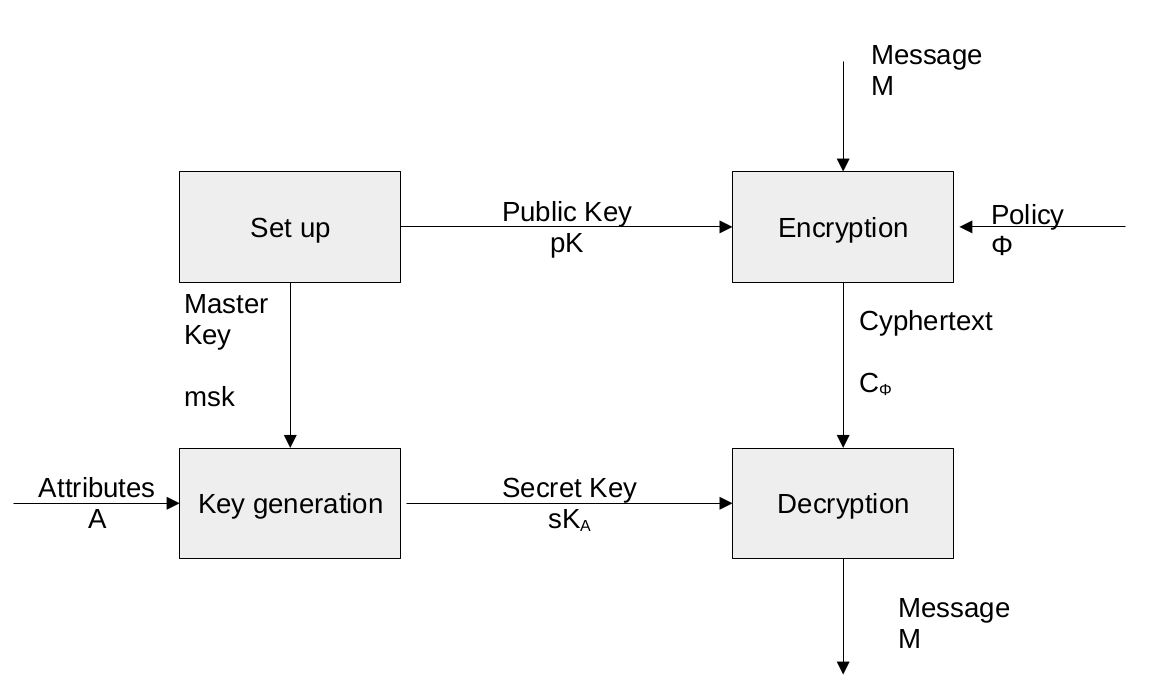}
    \caption{Ciphertext-policy Attribute-Based Encryption scheme \cite{ABE}}
    \label{fig:3}    
\end{figure}

\textbf{Ciphertext-policy attribute based encryption} (CP-ABE), as depicted in Figure \ref{fig:3}, consists in defining an access policy in the message itself. A user can decrypt a message if the attributes associated with his attributes satisfies the access policy embedded within the ciphertext. For example, if the whole set of attributes is defined as $\Omega = {A, B, C, D}$, a message is encrypted with the policy $\phi = (A \wedge B) \vee D$, a user provided with the attribute $D$ and its related secret can decrypt the message, whereas a user with the attributes ${A, B}$ cannot \cite{ABE}.\\

\begin{figure}[!h]
    \centering
    \includegraphics[scale=0.3]{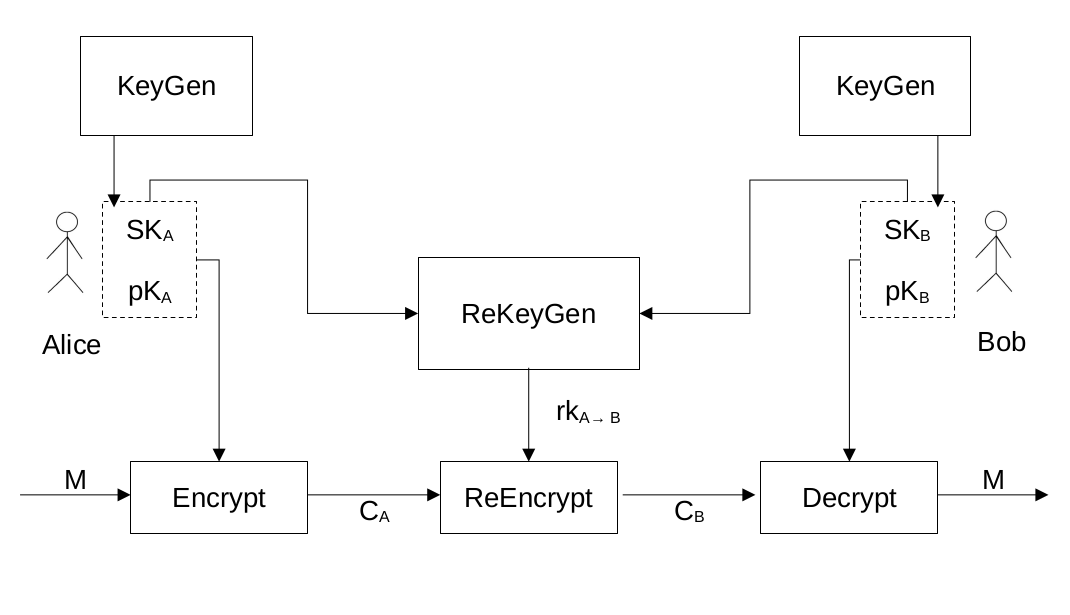}
    \caption{Ciphertext-policy Attribute-Based Proxy Re-Encryption \cite{report-crypto}}
    \label{fig:ROC2s}    
\end{figure}

In addition, we assume there is an existing ABE reencryption scheme for enabling a semi-trusted proxy to transform a ciphertext under an access policy $C_A$ to another ciphertext corresponding the same plaintext, but under another access policy $C_B$, as depicted in Figure \ref{fig:ROC2s}. The objective is that the proxy performs the reencryption operation blindly. That is, the proxy is not provided with attributes satisfying the policy of the ciphertext, and thus gets no information about the plaintext sent by Alice. Only entities satisfying $C_B$ like Bob can decrypt the ciphertext. Any scheme, like the  Ciphertext-Policy Attribute-Based Proxy Re-Encryption (CP-ABPRE) proposed in \cite{6630473}, can be integrated in our contribution.  
Attribute-Based Encryption is a recent mechanism and is not yet widely used, but many applications of it are possible in particular to broadcast a symmetric key. \\

\section{Our Security Architecture Combining SSI, IPFS and ABE}
\label{proposal}

In this section, we present our proposal. We propose a SSI infrastructure to manage the identity and access of involved entities to EHR. We encrypt EHR on two-levels using ABE and we store the encrypted records and cryptographic key material on IPFS.

A concrete reference scenario is ﬁrst introduced for illustrating the architecture in a healthcare context, followed by the technical design with the full interactions between the patient, the hospital, the doctor, the IPFS and the blockchain.
\\

We assume for our architecture that health institutions act as issuers for both doctors and patients. Health professionals (doctors, therapists ..) will have verifiable credentials issued to them by their institutions and the supervising authorities and bodies. As for patients, verifiable credentials attesting to their identity can be obtained from the competent authorities and other credentials attesting to their attributes can be obtained from different health authorities and institutions. These attributes along with ABE secret keys will be later used for ABE decryption (cf. Section 3.2).

Verifiable credentials and ABE secret keys are stored on users wallets.

\subsection{The Reference Scenario}

\begin{figure}[!h]
    \centering
    \includegraphics[scale=0.3]{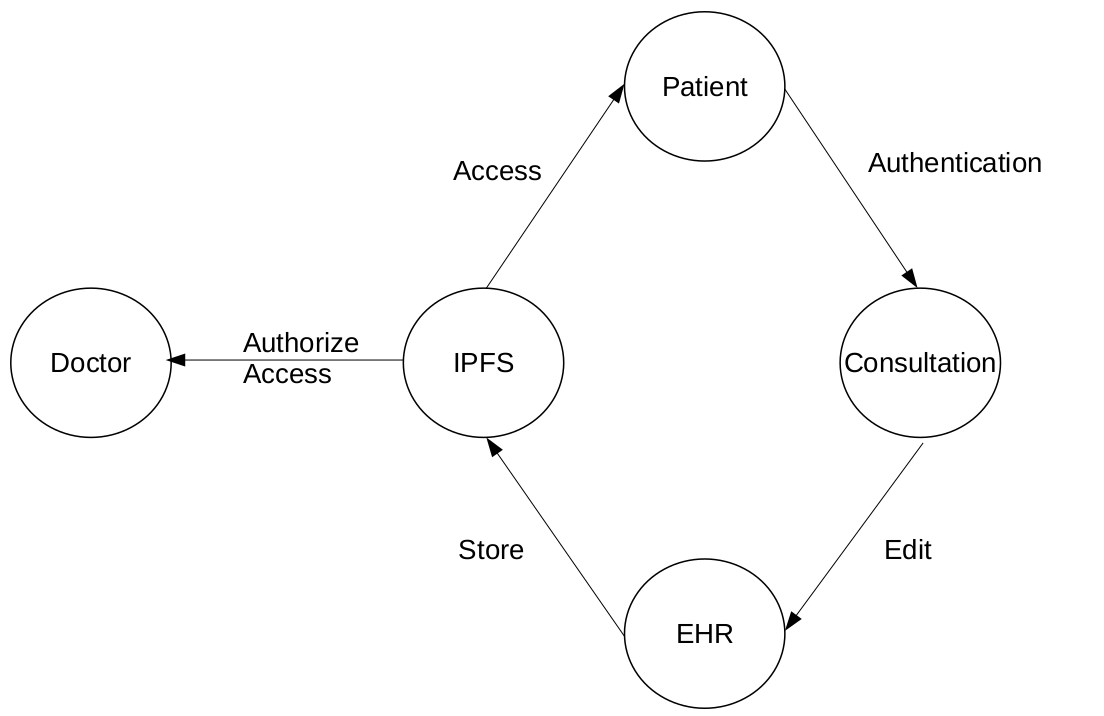}
    \caption{The reference scenario}
    \label{fig:ref_scena}    
\end{figure}

We describe the following healthcare scenario, depicted in Figure \ref{fig:ref_scena}. A patient arrives at a hospital or a healthcare institution, they perform a mutual authentication with the facility’s servers where verifiable credentials are presented from patient’s wallet. After successfully authenticating the patient’s identity, the medical appointment can take place.
EHR of the patient is edited after the appointment and is securely stored on the IPFS.
Patients can consult their records and consent to grant authorization to access these records to the medical staff.

\subsection{Solution Design}
\label{sec:Design}    
This subsection describes the underlying interactions between entities, for establishing a secure communication channel between two blockchain agents (patient and hospital, but also doctor and hospital) (cf. Section 4.2.1), for secure interactions and exchange of credentials, cryptographic material and EHR storage (cf. Section 4.2.2), for patients to modify access rights to their records (cf. Section 4.2.3), for letting an authorized doctor retrieve a patient’s record (cf. Section 4.2.4), and for enabling emergency access to records in case the patient is physically or mentally unable to give their consent (cf. Section 4.2.5). Section 4.2.6 is a full-picture summary of this subsection.

\subsubsection{Establishing a mutually authenticated channel between the hospital and the patient}
\label{sec:authentication}    

The two actors in this interaction, both patient and hospital, both have general DIDs and verified credentials attesting to their identities and natures, as we have supposed in the beginning of this section.
Through their agents, the hospital and the patient mutually authenticate themselves by exchanging credentials. After this mutual authentication, they create pair-wise DIDs special for this relationship, establishing a private secure channel.

\begin{figure}[!h]
    \centering
    \includegraphics[scale=0.3]{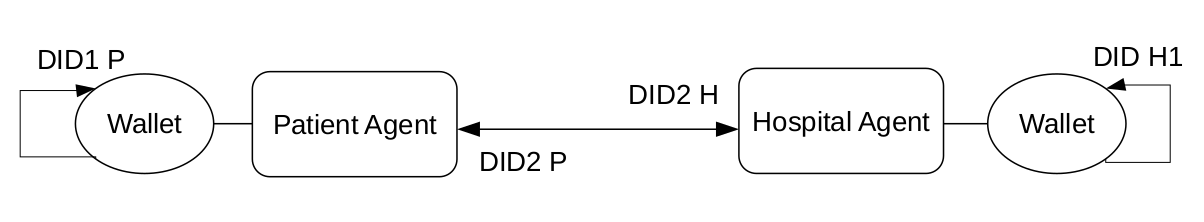}
    \caption{Establishing a mutually authenticated channel between the patient and the hospital by creating pair-wise DIDs for this specific relationship}
    \label{fig:comm_agents}    
\end{figure}

Figure \ref{fig:comm_agents} describes this channel creation.

\begin{figure}[!h]
    \centering
    \includegraphics[scale=0.3]{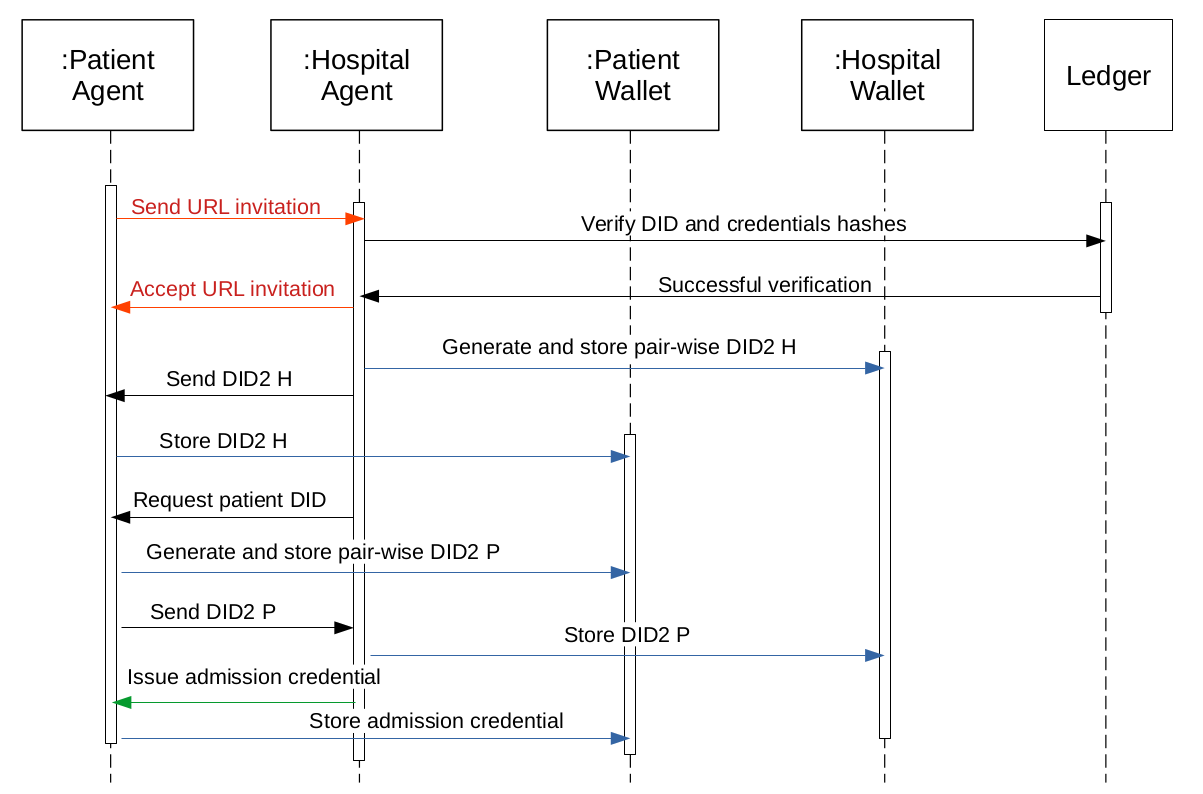}
    \caption{Establishing a secure mutually authenticated channel between the patient and the hospital}
    \label{fig:authent}    
\end{figure}

More precisely, when the patient ﬁrst arrives to the hospital, they undergo an initial authentication process, as explained in Figure \ref{fig:authent}. The patient’s agent sends an URL invitation with their global DID to the hospital’s agent along with any other verifiable credentials attesting to their identity. The hospital agent verifies the DID and associated credentials by referring to the blockchain Ledger. The patient does the same to achieve mutual authentication. 
Then the hospital generates a new pair-wise DID2 H speciﬁc to this particular patient, which is sent to the patient and stored in the hospital’s wallet. Similarly, the patient generates a new DID2 P speciﬁc to this communication channel with the hospital, which they send to the hospital and store on their wallet. The two agents have two pairwise DIDs for their own communication. Finally, as a last step, the hospital issues an admission credential for the patient which is stored on their wallet.

\subsubsection{Storing the Newly Issued Medical Record}
\label{sec:storing}    

\begin{figure}[!h]
    \centering
    \includegraphics[scale=0.3]{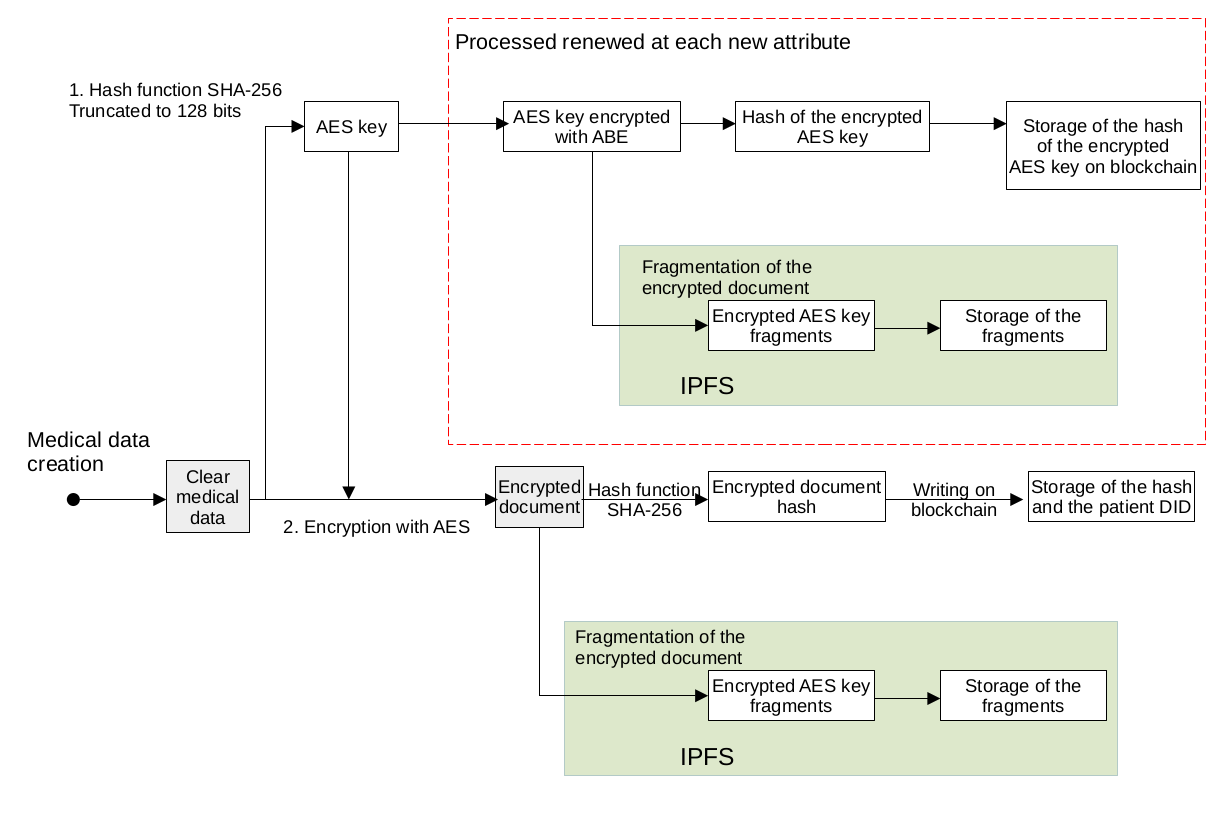}
    \caption{Encryption and storage of a newly issued Electronic Health Record or medical data}
    \label{fig:enc_stor}    
\end{figure}

The newly created EH record undergoes the various stages depicted in Figure \ref{fig:enc_stor}. First, the record is hashed using the SHA-256 hash function. It is then truncated to the ﬁrst 128 bits, which serve as the key for the symmetric AES encryption \cite{AES} of the record. Then, the encrypted record is stored in the IPFS storage capacity \cite{IPFS} and its hash value, known as the Content Identiﬁer (CI), is added to the blockchain with the patient’s DID. In the same way as for the secure storage
record, after the AES key is encrypted with the asymmetric ABE algorithm (cf. Section 3.2), the encrypted AES key is stored in the IPFS and its hash value is added to the blockchain with the hash of the encrypted document. The patient is informed about the CID of the newly stored record.

\subsubsection{Modifying the ABE Access Policy to Authorize Access to the EHR}  
\label{sec:modification}   
\begin{figure}[!h]
    \centering
        \includegraphics[scale=0.3]{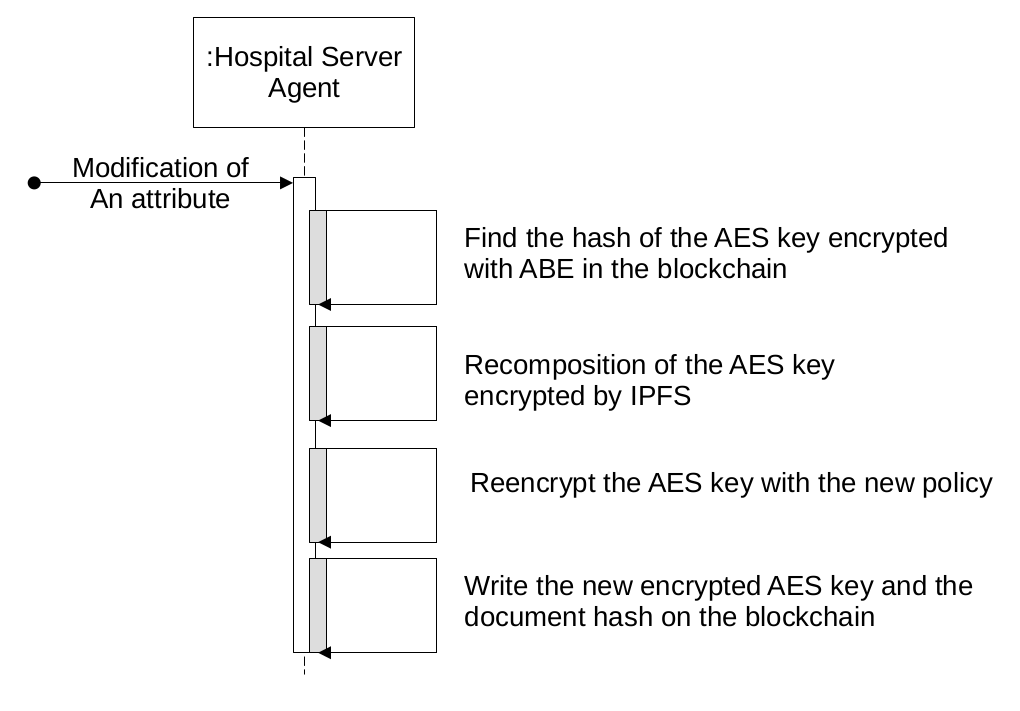}
    \caption{Elementary loop}
    \label{fig:elem_loop}    
\end{figure}

After a medical consultation, the EHR of a patient needs to be modified by a doctor, which means the patient have to consent to give access to their EHR to the doctor. The hospital agent is requested to perform the elementary loop depicted in Figure \ref{fig:elem_loop}. As a ﬁrst step, the agent has to recover the ABE encrypted AES key by ﬁrst retrieving the hash of the encrypted AES key from the blockchain thanks to the CID of the record, and then by getting the encrypted AES key value  from the IPFS. 

The resulting AES key is still encrypted with ABE. The hospital agent does not have the right attributes to decrypt the key and can only re-encrypt the AES key under a new policy including the newly considered attributes, by using the ABE re-encryption algorithm with the ABE re-encryption key provided by the patient (Section 3.2). 

As such, doctors satisfying the new policy related to the medical record can access to the record. The agent then has to update the resulting encrypted AES key into the system by writing its new hash value to the blockchain along with the same CID and the newly encrypted AES key into the IPFS. Note that multiple occurrences of the CID into the blockchain refers to the successive policy modification consented by the patient with regard to his CID record.

\subsubsection{Access to the EHR by healthcare professionals}
\label{sec:access}    
As soon as a health care professional is granted access to the patient’s EHR CID, they need to get the hash of the encrypted AES key from the blockchain using the CID value. They should also be able to retrieve the encrypted AES key value from the IPFS. 
A health care professional is able to decrypt the AES key, as their ABE attributes (stored within their wallet) satisfy the ABE access policy associated with the encrypted AES key (cf. Section 3.2), thanks to the ABE policy which has been modified by the patient (cf. Section 4.2.3) to enable them to decrypt the AES key. 
After decrypting the AES key, they can retrieve from IPFS the full encrypted CID medical record, and then can decrypt it with the AES key. Note that a health care professional who does not belong to the same service, I.e. his attributes do not satisfy the ABE policy, won’t be able to decrypt the AES key and so won’t have access to the EHR.

\subsubsection{Accountable Emergency Procedure}
\label{sec:emergency}  

In case a patient is unconscious and is not able to give their consent to authorize a health care professional to access their EHR, an emergency procedure can be used by the health care professionals.
Through an authenticated channel (cf. Section 4.2.1), the health care professional requests the Hospital Emergency Server (a blockchain agent as well), to modify the policy for letting them access to the patient’s EHR CID. The resulting emergency loop is different from the elementary loop of Section 4.2.3 as the hospital emergency server provided with attributes satisﬁes any
ABE policy associated to any record. It is thus able to fully decrypt the AES key, and to encrypt that AES key with the same policy increased with the attribute(s) owned by the requesting health care professional. 

The server then has to report to the blockchain the emergency procedure over the CID record requested by a requesting health care professional’s DID. In case of a later auditing procedure, it will be possible to evaluate whether that emergency record access was abusive or legitimate.

\begin{figure}[!h]
    \centering
    \includegraphics[scale=0.3]{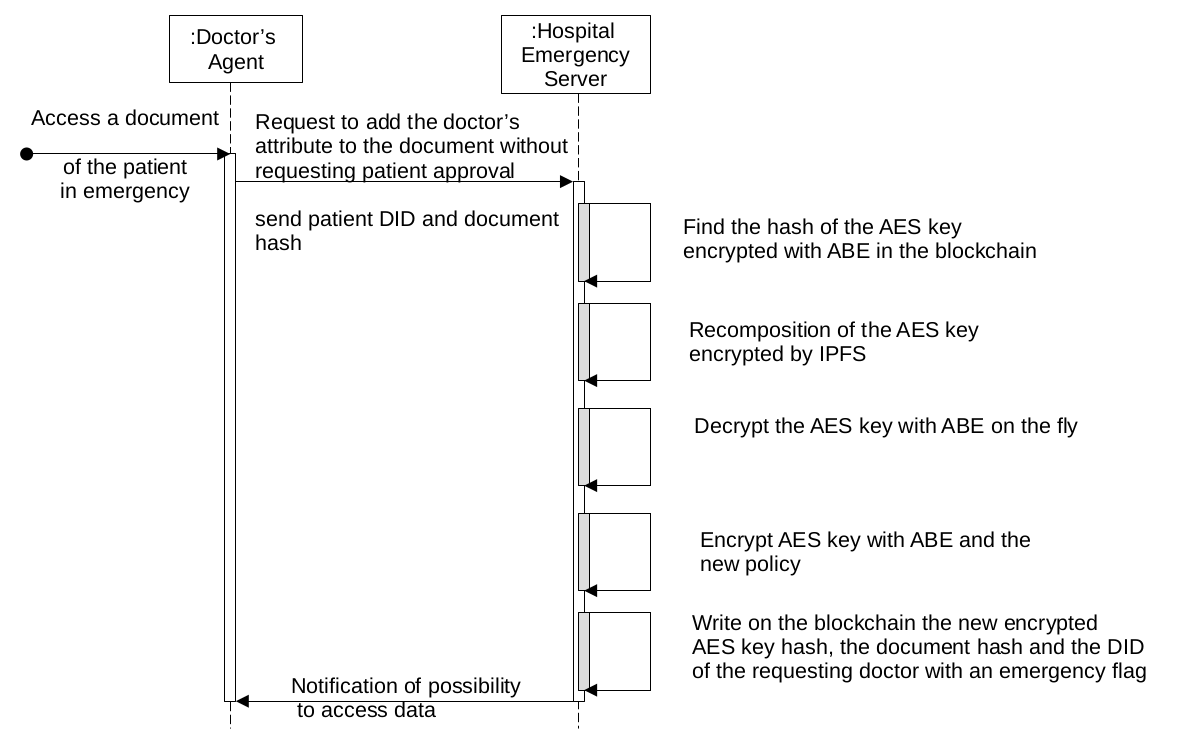}
    \caption{Emergency loop}
    \label{fig:emergency}    
\end{figure}

Figure  \ref{fig:emergency} depicts an emergency loop where a doctor requests access to the EHR of a patient incapable of giving their consent. The emergency server of the hospital performs an accountable procedure to grant access without consent of the patient however this access has to be later justified.

\subsubsection{The Full Picture of the Interactions between Patients and Medical Entities}
\label{sec:PatientInteraction}
\begin{figure}[!h]
    \centering
    \includegraphics[scale=0.3]{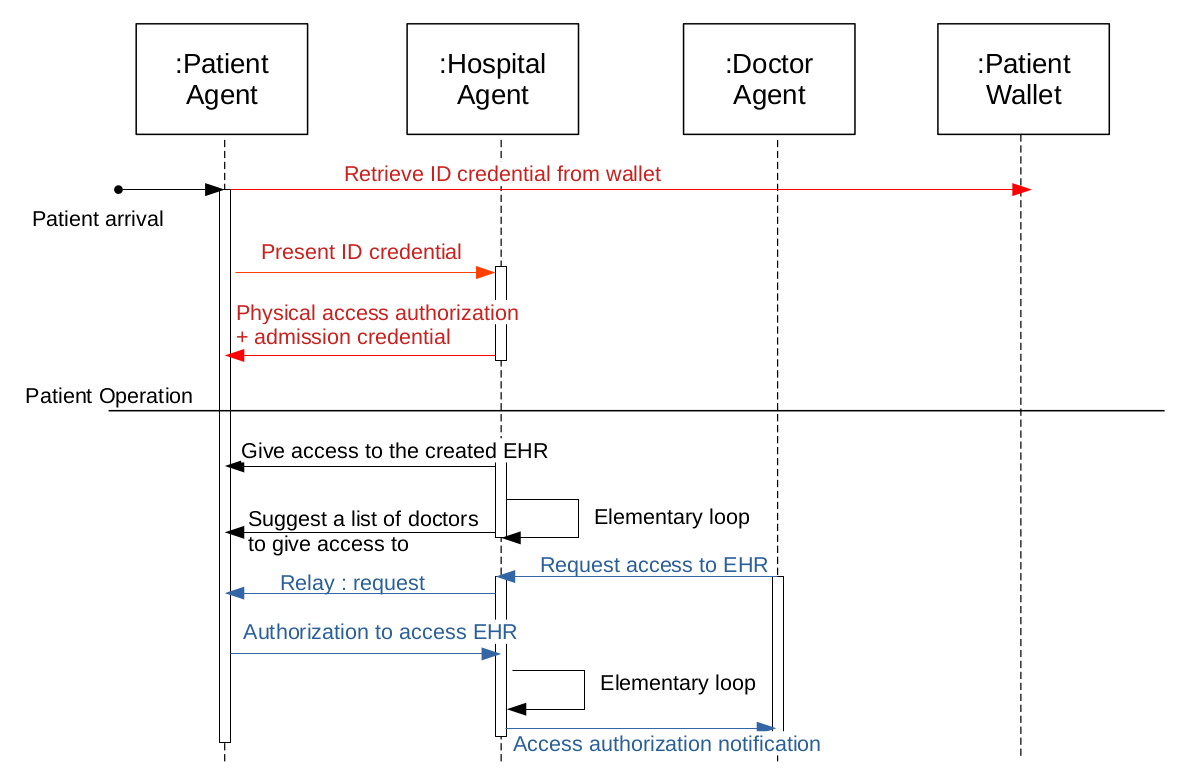}
    \caption{Patient interactions with medical entities}
    \label{fig:operating_room}    
\end{figure}

As depicted in Figure \ref{fig:operating_room}, at their arrival to the hospital, the patient retrieves an official ID credential from their wallet and present it to the hospital as part of the mutual authentication process  between the two agents. Upon establishing a mutually authenticated channel and creating a pairwise DID for this instance, an admission credential bound to the pairwise DID is issued by the hospital and sent to the patient who stores it in their wallet (cf. Section 4.2.1). The admission credential is used to grant access to the hospital spaces needed to complete the patient’s visit purpose.

Following the patient’s treatment, a medical record is created (X-Ray results, blood test results, analysis ..). The hospital performs the elementary loop (cf. Figure 9) to grant the record access to the patient. The hospital also suggests to the patient a list of doctors or health care professional that access might interest them or is needed for the patients treatment. A doctor or a health care professional needing access to a medical record makes a request to the hospital agent that later relays it to the patient’s agent. We assume that a mutually authenticated channel is similarly established between the doctor’s agent and the hospital’s agent.

When a patient grants access to a doctor upon receiving an access request from the hospital agent, the elementary loop is performed again to add the ABE attributes of the doctor to the access policy (procedure described in section 4.2.3). Subsequently, if the patient wishes to remove the access rights from a doctor, they make a request to the hospital agent that restarts the elementary loop and withdraws the doctor’s attributes.

\section{Analyzing the proposed Architecture}
\label{discussion}
Our proposed architecture has three important pillars: SSI as an identity model, Blockchain and IPFS as infrastructure and hybrid AES and ABE encryption as an access policy that ensures data privacy and confidentiality.
\subsection{SSI as an identity model}

The SSI model is suitable for healthcare use-cases since it is reasonable to give a patient full control over their identity and related data (EHR) in a context that allows for a portable identity and health data. This enables the patient to choose their healthcare providers and manage their EHR with a consented secure way.
Moreover, the SSI model allows for portable identifiers and identities, meaning that it is indeed interoperable and very scalable since SSI relies on a decentralized ledger and that the DID standard is ledger agnostic and globally unique.
These specifities mean that our architecture can work with any blockchain platform and that even when different healthcare networks choose different platforms, patients would still be able to move between these networks and move their data along with them.

On the downside, SSI requires secure user wallets and well designed blockchain agents. This sets higher standards and technological constraints on any SSI proposal since they are as good and strong as their wallet component. Moreover, user-awareness is needed to ensure that users can give their consent and manage their EHR data properly. 
\subsection{Blockchain and IPFS as infrastructure}
Blockchain and IPFS are two decentralized infrastructure technologies. Blockchain provides a decentralized public ledger to register DID identifiers and hashes of issued credentials, all along with access transactions like in the case of emergency access procedure described in Section 4.2.5.
Blockchain provides integrity and a permanent history that ensures accountability for any access to EHR data on the IPFS.

On the other hand, IPFS provides a decentralized storage platform for EHR data. IPFS overcomes the storage shortages of blockchain networks while maintaining the decentralization and public nature – meaning not owned and controlled by a single entity – all the same. Privacy however is not a built-in feature in IPFS systems since it is public and data is stored across scattered computers on the network. Meaning that privacy is a requirement that is added via anonymization of  data and cryptography to encrypt the stored data. IPFS follows a content based file system, meaning that searching for data on IPFS includes requesting content from the network, receiving a response from nodes showing different versions of the requested content. This content is encrypted and possibly signed digitally, and the hashing ensures it’s integrity (IPFS is based on DHT).

Relying on a blockchain and IPFS ensures mitigation against Denial of Service (DoS) due to their decentralized nature and the distributed architecture of the infrastructure.
However, decentralization comes at the cost of performance and throughput of the services since it takes longer to read and write data on such infrastructure, combined with an overhead of cryptography and hash functions used on both the blockchain and IPFS.
\subsection{Hybrid AES and ABE encryption for access control}
Encryption is more of a requirement than a luxury in our proposal. Storing data on IPFS, as specified in Section 5.2, we need encryption to ensure data privacy on a public or a consortium storage network.
However, this encryption is also used for access control, and more specifically, since IPFS is a content based file system, we use ABE encryption for a content-based access control.
We propose a AES and ABE hybrid encryption for our architecture, this means that a modified policy will only require the AES key to be ABE encrypted, making it more efficient.

ABE in our proposal provides a content-based access control, ensure mitigation against medical records leakage and unauthorized access. Granting access and revoking it is described in Section 4.2.3, however, one must be aware that a healthcare professional which access is removed is still able to decrypt a medical record previously stored on their hard drive.
\section{Conclusions}
\label{conclusion}

Managing medical data and EHR is an important factor for the success of any digital health application and services. The design of new architectures capable of resisting newer types of attacks is essential for the adoption of digital health services.

 Moreover, these architectures should be based on secure infrastructures that can guarantee data integrity and non-repudiation. As medical data and EHR are private and personal data, these architectures should take into account the privacy of the data and grant full control over it to the data subjects: the patients.

In this paper, we have proposed a self-sovereign healthcare architecture with an original fully distributed content-based access control. It combines several concepts and technologies that ﬁt well with the spirit of the decentralized solution: blockchain, self-sovereign identity (SSI), hybrid encryption including attribute-based encryption and distributed storage system. 

This architecture, once deployed with a smooth enrollment procedure, has the advantage of being scalable throughout hospitals and any medical institutions, with strengthened security
thanks to high resistance against denial of service attacks and data leakage by using encryption and making use of the high availability and decentralization of blockchain technology.

\section*{Acknowledgements}
Special thanks to Arthur Jovart for his technical support and for the chair BOPA (BOPA for augmented operating room) jointly mounted between Institut Mines-Telecom and AP-HP public hospitals of Paris for fruitful discussions.

\section*{Funding statement}This paper is partly supported by the Chair Values and Policies of Personal Information of Institut Mines-Telecom.

\end{document}